\documentclass[aps, prl, twocolumn, nofootinbib, floatfix, superscriptaddress, preprintnumbers]{revtex4-1}

\usepackage{comment}
\usepackage{graphicx}
\usepackage{dcolumn}
\usepackage{bm}
\usepackage{subfigure}
\usepackage{epsfig}
\usepackage{amsmath}
\usepackage{amsfonts}
\usepackage{mathrsfs}
\usepackage{amssymb}
\usepackage{nccmath}
\usepackage{amssymb,amsmath,amsfonts}
\usepackage[scr=boondoxo]{mathalfa}
\usepackage{xfrac}
\usepackage{color}
\usepackage{tikz}
\usepackage[T1]{fontenc}
\usepackage[utf8]{inputenc}
\usepackage{ulem}

\usepackage{tcolorbox}
\usepackage{hyperref}
\usepackage{booktabs}
\hypersetup{colorlinks=true, citecolor=red, linkcolor=blue, urlcolor=blue}

\definecolor{rossos}{cmyk}{0,1,1,0.55}
\definecolor{bluscuro}{rgb}{0.15, 0.2, .85}
\definecolor{bluchiaro}{cmyk}{1,.3,0.,0.1}

\newcommand{\be}{\begin{equation}}

\begin{document}

\preprint{YITP-26-58, KEK-QUP-2026-0010, KEK-TH-2844} 

\rightline{}
\title{Universal Suppression of Gravitational Waves from Black Hole Evaporation Dynamics}

\author{Xin-Chen He}
\email{xinchenhe@mail.ustc.edu.cn}
\affiliation{Department of Astronomy, School of Physical Sciences, University of Science and Technology of China, Hefei 230026, China}
\affiliation{CAS Key Laboratory for Researches in Galaxies and Cosmology, School of Astronomy and Space Science, University of Science and Technology of China, Hefei, Anhui 230026, China}
\affiliation{Asia Pacific Center for Theoretical Physics, Pohang 37673, Korea}

\author{Xiao-Han Ma}
\email{mxh171554@mail.ustc.edu.cn}
\affiliation{Department of Astronomy, School of Physical Sciences, University of Science and Technology of China, Hefei 230026, China}
\affiliation{CAS Key Laboratory for Researches in Galaxies and Cosmology, School of Astronomy and Space Science, University of Science and Technology of China, Hefei, Anhui 230026, China}
\affiliation{Asia Pacific Center for Theoretical Physics, Pohang 37673, Korea}
\affiliation{Kavli Institute for the Physics and Mathematics of the Universe (WPI), UTIAS The University of Tokyo, Kashiwa, Chiba 277-8583, Japan}

\author{Misao Sasaki}
\email{misao.sasaki@apctp.org}
\affiliation{Asia Pacific Center for Theoretical Physics, Pohang 37673, Korea}
\affiliation{Pohang University of Science and Technology (POSTECH)
77 Cheongam-Ro. Nam-Gu, 790-784 Pohang, South Korea}
\affiliation{Kavli Institute for the Physics and Mathematics of the Universe (WPI), UTIAS The University of Tokyo, Kashiwa, Chiba 277-8583, Japan}
\affiliation{Center for Gravitational Physics, Yukawa Institute for Theoretical Physics, Kyoto University, Kyoto 606-8502, Japan}
\affiliation{Leung Center for Cosmology and Particle Astrophysics,
National Taiwan University, Taipei 10617, Taiwan}

\author{Volodymyr Takhistov}
\email{vtakhist@post.kek.jp}
\affiliation{International Center for Quantum-field Measurement Systems for Studies of the Universe and Particles (QUP), KEK, 1-1 Oho, Tsukuba, Ibaraki 305-0801, Japan}
\affiliation{Theory Center, Institute of Particle and Nuclear Studies, High Energy Accelerator Research Organization (KEK), Tsukuba 305-0801, Japan}
\affiliation{Graduate University for Advanced Studies (SOKENDAI), 1-1 Oho, Tsukuba, Ibaraki 305-0801, Japan}
\affiliation{Kavli Institute for the Physics and Mathematics of the Universe (WPI), UTIAS The University of Tokyo, Kashiwa, Chiba 277-8583, Japan}


\begin{abstract}
Evaporating black holes can leave distinct imprints on gravitational wave (GW) backgrounds.
We show that black hole populations with finite width mass distributions exhibit a universal late
time evolution governed by the evaporation dynamics rather than the details of the initial mass
distribution, leading to a characteristic power law suppression of the induced GWs. We demonstrate
this for a broad class of mass functions in primordial black hole (PBH) scenarios featuring an early
Universe matter-dominated era, and identify the suppression of PBH-induced GWs found for critical
collapse distributions as a manifestation of this general phenomenon. Our results establish a direct
connection between the asymptotic GW spectrum and the underlying law of black hole evaporation.
\end{abstract}

\maketitle

{\it{Introduction.}}-- 
Hawking radiation from black holes~\cite{Hawking:1974rv,Hawking:1975vcx} is a fundamental prediction of quantum field theory in curved spacetime, yet no direct observational confirmation exists to date. Primordial black holes (PBHs)~\cite{Zeldovich:1967lct,Hawking:1971ei,Carr:1974nx}, black holes potentially formed in the early Universe  such as from the gravitational collapse of large density fluctuations, offer a distinct cosmological testbed in which black hole evaporation dynamics can leave observable imprints. If a sufficiently abundant population of light PBHs temporarily dominates the energy density before evaporating~\cite{Carr:1975qj,Khlopov:1980mg,Carr:2020gox}, the resulting reheating transition to a radiation-dominated (RD) Universe sources a stochastic gravitational-wave (GW) background through scalar-induced GWs~\cite{Ananda:2006af,Baumann:2007zm,Kohri:2018awv,Domenech:2021ztg}. 

Many analyses have considered such induced GWs in the idealized limit where all black holes evaporate simultaneously, corresponding to a monochromatic mass distribution. 
This instantaneous evaporation produces an abrupt transition from the early matter-dominance (eMD) to the radiation-dominance (RD), leading to a characteristic high-frequency enhancement of the GW spectrum~(e.g.,~\cite{Inomata:2019ivs, Domenech:2020ssp, Papanikolaou:2020qtd}). 
Several studies have utilized this prominent signal as a novel probe for the spatial distribution of ultra-light PBHs and the associated primordial non-Gaussianity~\cite{Papanikolaou:2024kjb,He:2024luf}.
Beyond this idealized monochromatic approximation, the physical relevance of an extended PBH mass distribution has also been recognized~\cite{Domenech:2021wkk,Papanikolaou:2022chm}. 
However, it was generally anticipated that as the mass distribution broadens, the gravitational potential suppression would merely exhibit a progressive deviation from the monochromatic scaling $\mathscr{S}_{\Phi}\propto k^{-1/3}$
~\cite{Inomata:2020lmk,Pearce:2023kxp,Pearce:2025ywc}.  
Recently, a steeper suppression scaling  $\mathscr{S}_{\Phi}\propto k^{-4/3}$  was reported for critical gravitational collapse distributions~\cite{Gouttenoire:2026pmp,Gouttenoire:2026pem}. This raises the question of whether the suppression is specific to this mass function or is instead a universal feature of black hole evaporation dynamics.

 \begin{figure*}[t] 
    \includegraphics[width=0.48\textwidth] {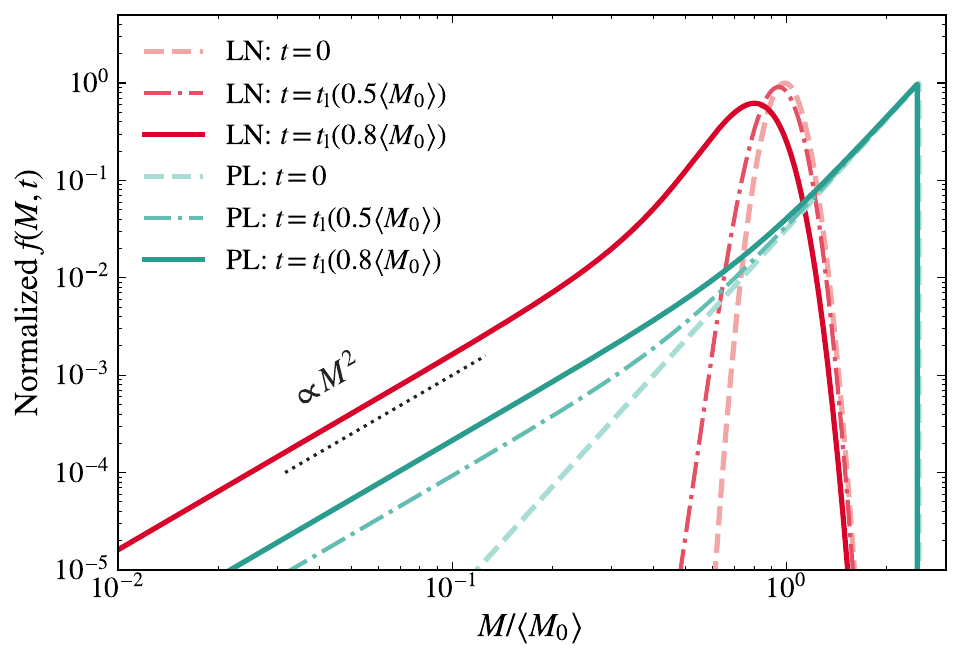} 
    \hspace{1em}
    \includegraphics[width=0.48\textwidth]{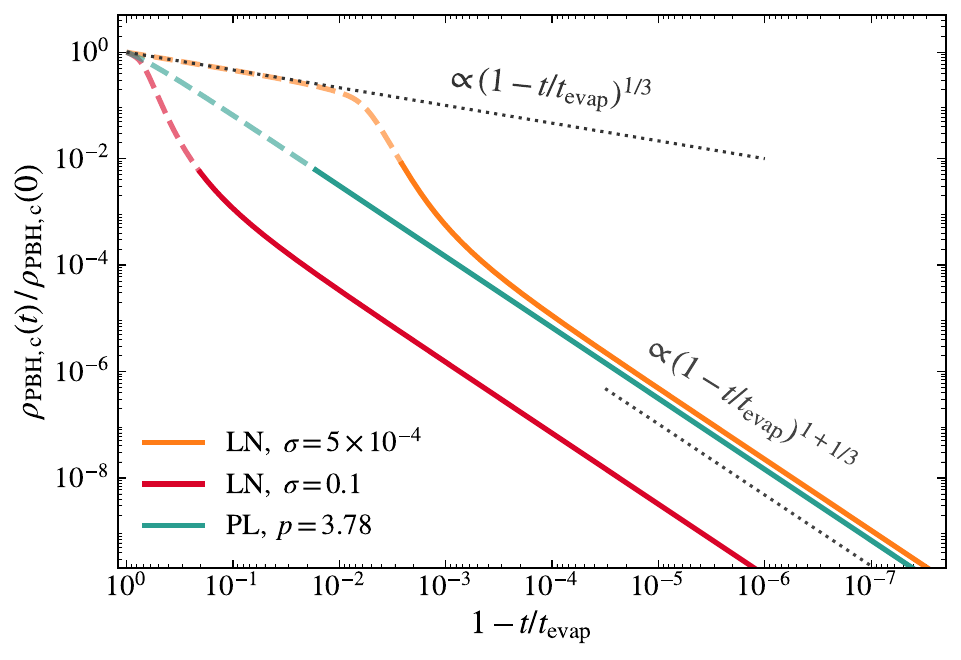} 
\caption{
Universal endpoint evolution of evaporating PBH mass functions. [Left] Instantaneous mass distributions for representative log-normal (LN) with $\sigma = 0.1$ and power law (PL) with $p = 3.78$ initial distributions (with $\langle M_0 \rangle$ denoting the initial mean mass) considering Hawking evaporation ($\alpha = 2$), displaying the universal endpoint tail $f(M,t)\propto M^2$. [Right] Comoving PBH energy density near complete evaporation. Finite-width distributions asymptote to $\rho_{\rm PBH,c}\propto (1 - t/t_{\rm evap})^{1 + 1/(1+\alpha)}$, while the monochromatic case follows the singular scaling $(1 - t/t_{\rm evap})^{1/(1+\alpha)}$.
}
\label{fig:scaling_rho_pbh}
\end{figure*} 

In this Letter, we show that the suppression of the gravitational potential associated with evaporating black hole populations is a universal consequence of the evaporation law, rather than of the detailed form of the initial mass distribution. 
For any finite-width mass distribution with an effective maximum mass, the number of surviving black holes decreases continuously as evaporation nears completion, in contrast to the idealized monochromatic limit where all black holes vanish simultaneously.

For Hawking evaporation, we show that the universal black hole evaporation endpoint dynamics reproduce  across a broad class of mass functions  the suppression found in Ref.~\cite{Gouttenoire:2026pmp,Gouttenoire:2026pem}. We identify its origin as the combined effect of evaporation mass loss and continuous depletion of the surviving population, not as a special property of the critical collapse mass function. More broadly, the resulting high frequency asymptotic behavior of induced GWs encodes the underlying evaporation dynamics, connecting cosmological GW signatures to black hole microphysics.

\textit{Dynamics of evaporating compact objects.}\label{sec: Evaporation Dynamics}-- 
 We consider a comoving volume containing a population of evaporating compact  objects, focusing on PBHs. Let $f(M,t)$ denote the comoving mass distribution, such that $f(M,t) dM$ is the comoving number density of objects with masses in the interval $[M,M+dM]$. 
It obeys a continuity equation  
\begin{equation}\label{eq:continuity}
    \partial_t f(M,t)+\partial_M \left[\dot M f(M,t)\right]=0,
\end{equation}
where the overdot denotes a derivative with respect to cosmic time $t$, so that
$\dot M = dM/dt$ is the mass loss rate of an individual object.

Eq.~\eqref{eq:continuity} is general and applies whenever the object population evolution can be described by an effective mass loss law depending just on $M$. 
Thus, this describes non-rotating, non-charged PBHs as well as scenarios in which other parameters such as spin cease to play a prominent role in the endpoint evolution.

The continuity equation can  be solved analytically, as described in Supplemental Material, giving  
\begin{equation}\label{eq:solution_of_f_main}
    f(M,t) = f_0\bigl(M_0(M,t)\bigr)
    \frac{
        \dot M\bigl(M_0(M,t)\bigr)
    }{
        \dot M(M)
    }~,
\end{equation}
where $f_0(M) =  f(M,0)$ is the initial mass function, and $M_0(M,t)$ denotes the initial mass corresponding to a PBH with mass $M$ at time $t$.

For Hawking evaporation, the mass loss takes the form   $\dot M=- \kappa_H/M^2$, where prefactor $  \kappa_H$  depends on the number of particle species emitted and on greybody factors. 
Since the endpoint scaling derived below depends only on the power  law form of the mass loss rate, to illustrate the universal scaling behavior it is useful to consider a more general characterization of the evaporation law,  
\begin{equation}\label{eq:Evaporation law}
    \frac{dM}{dt} = -\frac{\kappa}{M^\alpha}~.
\end{equation}
We restrict our analysis to $\kappa>0$ and $\alpha>-1$, with Hawking evaporation corresponding to the case of $\alpha=2$. This generalized form allows the asymptotic scaling to be related directly to possible modifications of the evaporation law.
The same framework can be applied to other gradually depleting relic populations, such as Q-balls. In the case of gauge-mediated Q-balls, $M\propto Q^{3/4}$ and $dQ/dt\propto -Q^{-1/4}$ \cite{Kusenko:1997si}, which gives $\alpha=2/3$.

From Eq.~\eqref{eq:Evaporation law}, 
a black hole of initial mass $M_0$  evolves as $M(t)=M_0\left(1-t/t_{\rm l}\right)^{1/(\alpha+1)}$ for $t<t_{\rm l}$, where $t_{\rm l}=M_0^{\alpha+1}/[(\alpha+1)\kappa]$ is its lifetime. 
Using Eq.~\eqref{eq:solution_of_f_main} one has
\begin{equation}\label{eq:fMalpha}
    f(M,t)
    =
    \left(
    \frac{M}{M_0}
    \right)^\alpha
    f_0(M_0),
\end{equation}
where
$M_0=\left[M^{\alpha+1}+(\alpha+1)\kappa t\right]^{1/(\alpha+1)}$
is the initial mass corresponding to a mass $M$ at time $t$.

For any fixed time $t$ after the evaporation of the lightest PBHs and in the  low mass regime $M^{\alpha+1}\ll(\alpha+1)\kappa t$, the corresponding initial mass becomes independent of $M$, namely $M_0(M,t)\simeq[(\alpha+1)\kappa t]^{1/(\alpha+1)}$. Therefore, Eq.~\eqref{eq:fMalpha} gives a universal scaling $f(M,t)\propto M^\alpha$, independent of the initial mass function.
Physically, for $\alpha>0$, lighter PBHs evaporate faster. Hence, any 
finite width PBH population is dynamically driven toward the same low mass distribution.

\begin{figure*}[t]
    \includegraphics[width=0.48\textwidth]{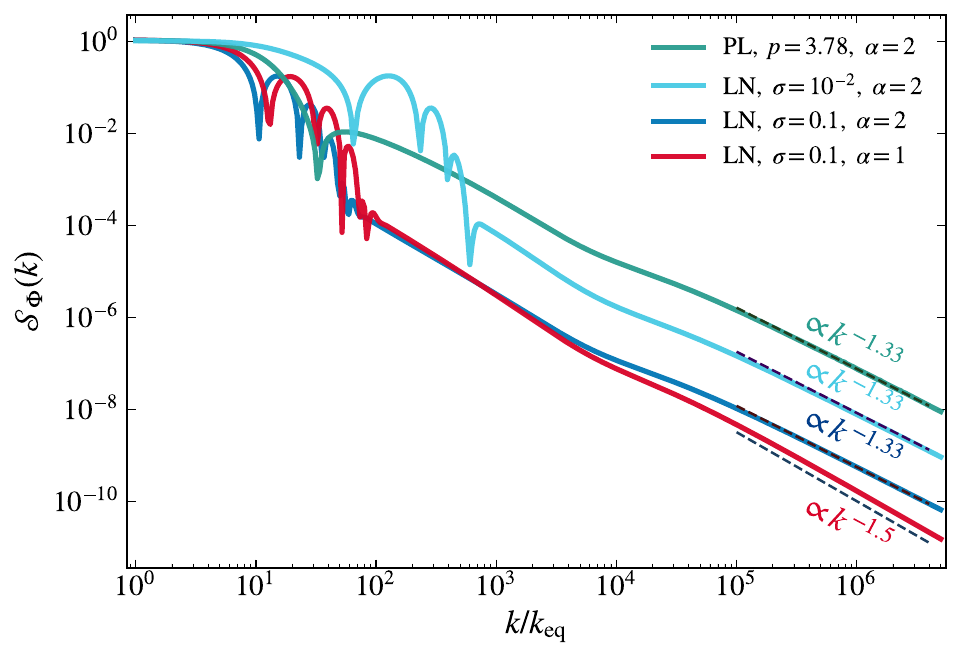}
    \includegraphics[width=0.48\textwidth]{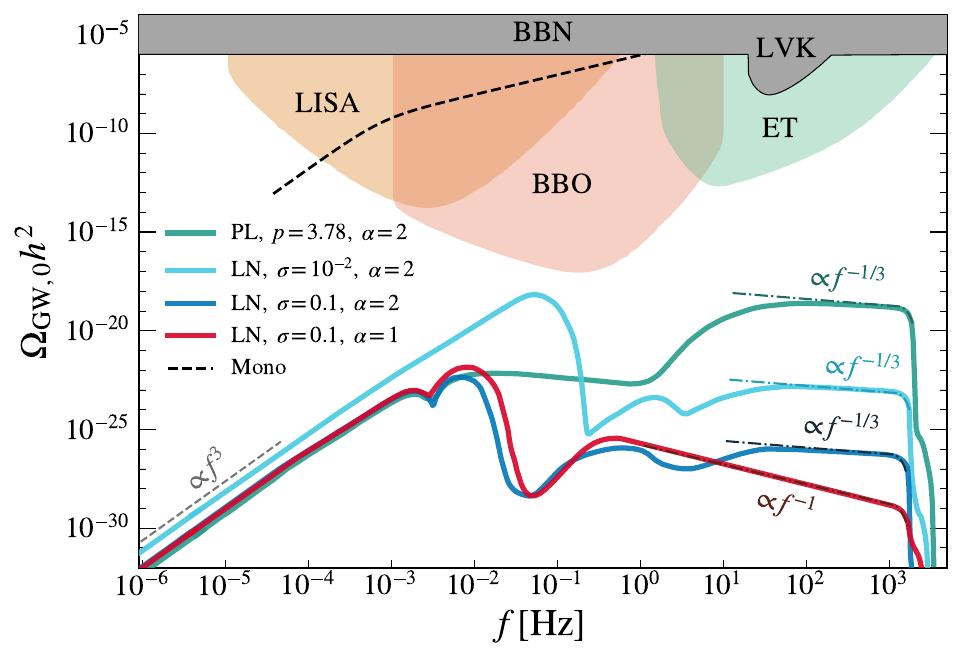}
\caption{
Universal suppression and induced GWs. [Left] Suppression factor $\mathscr{S}_{\Phi}(k)$ for representative mass functions, showing the high-$k$ scaling
$\mathscr{S}_{\Phi} \propto k^{-(\alpha+2)/(\alpha+1)}$.
For Hawking evaporation ($\alpha=2$), this gives $\mathscr{S}_\Phi \propto k^{-4/3}$.
[Right] Scalar induced GW spectra, illustrating the suppressed high frequency tail for finite width evaporation histories. All curves assume an initial PBH fraction $\varrho_{\rm PBH}/\varrho_{\rm tot}=2\times10^{-5}$ considering a reference PBH mass (initial central mass for LN, and cutoff mass for PL) of $10^4~\text{g}$. 
For this mass, the PBH lifetime of LN with
 $\alpha = 1$ is normalized to the PBH lifetime with $\alpha = 2$. The  monochromatic prediction (black dashed), evaluated at the same mass scale and initial fraction for comparison, is already restricted by BBN constraints~\cite{Domenech:2020ssp}. Stochastic GW background constraints from LVK~\cite{LIGOScientific:2025bgj} and projected sensitivities of LISA~\cite{LISA:2017pwj,Karnesis:2022vdp}, BBO~\cite{Harry:2006fi}  and ET~\cite{Maggiore:2019uih} are also shown.} 
\label{fig:OmegaGW}
\end{figure*}

The comoving energy density of PBH  population is \begin{equation}\label{eq:rho_PBH_comoving} \rho_{\rm PBH,c}(t) = \int_0^{M_{\rm max}(t)} M f(M,t) dM , \end{equation} where $M_{\rm max}(t)$ is the largest surviving mass at time $t$, corresponding to initial mass $M_{\rm max,0}$. The PBH population thus completely evaporates at $t_{\rm evap}=M_{\rm max,0}^{\alpha+1}/[(\alpha+1)\kappa]$. For realistic distributions with exponentially suppressed high mass tails, such as a lognormal, $M_{\rm max,0}$ is the effective maximum mass relevant within the finite cosmological volume or observable range.

Near complete evaporation time $t_{\rm evap}$, the upper mass boundary of the surviving population is $M_{\rm max}(t)=[(\alpha+1)\kappa(t_{\rm evap}-t)]^{1/(\alpha+1)}$. The integral in Eq.~\eqref{eq:rho_PBH_comoving} is determined by the low mass tail of the distribution $f(M,t)$. Since $f(M,t)\propto M^{\alpha}$, the late time scaling yields 
\begin{equation}\label{eq:rhoPBH_scaling_main} \rho_{\rm PBH,c}(t) \propto M_{\rm max}^{\alpha+2}(t) \propto (t_{\rm evap}-t)^{1 +\frac{1}{\alpha+1}} .
\end{equation} 
The comoving number density scales as $n_{\rm PBH,c}(t)\propto (t_{\rm evap}-t)$, as shown in the Supplemental Material.

The implications of Eq.~\eqref{eq:rhoPBH_scaling_main} are further highlighted by contrast with a
monochromatic population. There, all PBHs evaporate simultaneously and $n_{\rm PBH,c}(t)$ remains
constant until $t_{\rm evap}$ and then drops discontinuously to zero. Hence, one would have
$\rho_{\rm PBH,c}\propto M(t)\propto(t_{\rm evap}-t)^{1/(\alpha+1)}$.
Instead, a finite width PBH mass distribution continuously depletes with $n_{\rm PBH,c}(t)\propto(t_{\rm evap}-t)$, contributing an additional  power to
$\rho_{\rm PBH,c}\sim n_{\rm PBH,c}(t) M(t)$ and yielding Eq.~\eqref{eq:rhoPBH_scaling_main}.

Fig.~\ref{fig:scaling_rho_pbh} shows the mass function evolution
for Hawking evaporation (i.e. $\alpha=2$), where our general solution naturally reproduces scaling identified in Refs.~\cite{Carr:2016hva, Cai:2021fgm}. As representative examples, we consider
a log-normal (LN) PBH distribution with a peak width of $\sigma$ arising in various inflationary scenarios~\cite{Dolgov:1992pu,Kannike:2017bxn}, and a power law (PL) PBH distribution capturing
the low mass tail of critical collapse~\cite{Choptuik:1992jv,Evans:1994pj,Niemeyer:1999ak} with a PL slope of $p=1+1/\gamma_M\simeq3.78$,
following Ref.~\cite{Gouttenoire:2026pem,Gouttenoire:2026pmp}. Despite distinct
initial profiles, both evolve to the same universal $f\propto M^\alpha=M^2$
at late times. 
In Fig.~\ref{fig:scaling_rho_pbh}, we also show the evolution of $\rho_{\rm PBH,c}$
that  confirms  the additional power $\propto(t_{\rm evap}-t)$.

{\it{Suppression of gravitational potential.}}
\label{sec:PBH_matter_PS_non_Gaussian}--
The universal late time scaling of $\rho_{\rm PBH,c}$ has significant implications
whenever PBHs dominate the early Universe.
For the PBHs to evaporate before Big Bang nucleosynthesis (BBN), their mass must satisfy
$M_{\rm PBH}\lesssim10^{8}~{\rm g}$~\cite{Carr:2020gox}.  eMD eras can also arise from other non-relativistic relic objects, such as Q-balls~\cite{Chiba:2009zu} that decay gradually and whose induced GWs are suppressed, while oscillons~\cite{Lozanov:2022yoy,Lozanov:2023aez,Lozanov:2023knf,Lozanov:2023rcd} need not be, depending on the scenario.

On scales sufficiently larger than the mean PBH separation, its population can be treated as a pressureless fluid carrying primordial isocurvature
perturbations~\cite{Papanikolaou:2020qtd,Domenech:2021wkk}.
Its coupled background evolution
with the radiation produced by evaporation obeys the continuity equations
\begin{align}
    \dot{\varrho}_{\rm PBH} + 3H\varrho_{\rm PBH} &= -\Gamma(t) \varrho_{\rm PBH} , \\
    \dot{\varrho}_{\rm r} + 4H\varrho_{\rm r} &=  \Gamma(t) \varrho_{\rm PBH} ,
\end{align}
where $\varrho_{\rm PBH}$ and $\varrho_{\rm r}$ are physical PBH and radiation densities. 
The expansion rate $H =\dot a/a$ 
is set by
$3M_{\rm Pl}^2H^2=\varrho_{\rm PBH}+\varrho_{\rm r}$ with $M_{\rm Pl} = 1/\sqrt{8\pi G}$ being the reduced Planck mass and $G$ the gravitational constant. 
The collective mass loss is
encoded in $\Gamma(t) =-\dot\rho_{\rm PBH,c}/\rho_{\rm PBH,c}$, defined through
the comoving density so as to isolate the cosmological dilution effect.

Turning to perturbations, the induced GWs are sourced at second order by the gravitational potential $\Phi$, which on sub-horizon scales at the PBH-dominated stage obeys the Poisson equation $k^2\Phi\sim a^2\varrho_{\rm PBH}\delta_{\rm PBH}/M_{\rm Pl}^2$.
Combining this with $\varrho_{\rm PBH}=a^{-3}\rho_{\rm PBH,c}$ gives
$\Phi\propto\rho_{\rm PBH,c}$. Since $\rho_{\rm PBH,c}$ is nearly constant before
significant evaporation, $\Phi$ remains at a value $\Phi_k^{\rm eMD}$
throughout the eMD era. Once the matter component decays, its perturbations are
inherited by the radiation fluid, and hence by the induced GWs.
The GW signal is therefore highly sensitive to how $\Phi$ evolves across the transition.

A key effect arises  during the final stages of evaporation, where the evaporation rate
$\Gamma$ becomes higher than $H$. 
From the scaling of Eq.~\eqref{eq:rhoPBH_scaling_main} as $t \to t_{\rm evap}$, $\Gamma$ develops a  pole 
$\Gamma(t)=\frac{\alpha+2}{\alpha+1}(t_{\rm evap}-t)^{-1}$.
Over this short interval
$a$ and $\delta_{\rm PBH}$ vary only mildly  so $\Phi\propto\rho_{\rm PBH,c}$
continues to hold and the suppression of $\Phi$ is controlled by the decay of
$\rho_{\rm PBH,c}$. The potential tracks $\rho_{\rm PBH,c}$ until a decoupling time
$t_{\rm dec}(k)$ set by $\Gamma\simeq k/a$, beyond which the radiation perturbation
takes over and the suppression saturates. This defines the net suppression factor 
$\mathscr{S}_\Phi(k)=\Phi_k(t_{\rm evap})/\Phi_k^{\rm eMD}$. For large $k$,
$t_{\rm dec}$ lies close to $t_{\rm evap}$, where $a\simeq a_{\rm evap}$ and
$(t_{\rm evap}-t_{\rm dec})\propto k^{-1}$. With
$\mathscr{S}_\Phi(k)\propto\rho_{\rm PBH,c}(t_{\rm dec})$ and
Eq.~\eqref{eq:rhoPBH_scaling_main} this gives
\begin{equation}\label{eq:S_Phi_main}
    \mathscr{S}_\Phi(k)\propto k^{- \frac{\alpha+2}{\alpha+1}}=k^{-1-\frac{1}{\alpha+1}} ,
\end{equation}
while the monochromatic case gives $\mathscr{S}_\Phi^{\rm mono}\propto
k^{-1/(\alpha+1)}$. The extra factor $k^{-1}$ is independent of $\alpha$ and arises
from the universal linear depletion $n_{\rm PBH,c}\propto(t_{\rm evap}-t)$ of the
surviving population, absent when all PBHs evaporate at once. We display in Fig.~\ref{fig:OmegaGW} these results. 
Since 
$\Omega_{\rm GW}\sim\langle\Phi^4\rangle\propto
\mathscr{S}_\Phi^4$, this gives an additional $k^{-4}$ suppression of the high $k$ frequency spectrum relative to the monochromatic case, with a qualitatively different spectral shape.

{\it{Induced gravitational waves.}}\label{sec:induced_GW}--
Induced GWs sourced at
the eMD-RD transition have been studied with a focus on both
adiabatic~\cite{Inomata:2019ivs,Inomata:2020lmk} and
isocurvature~\cite{Papanikolaou:2020qtd,Domenech:2021wkk} initial perturbations. Since the suppression of Eq.~\eqref{eq:S_Phi_main} originates from the transition
itself, it applies regardless of the perturbation type. To minimize model assumptions, here we focus on the unavoidable isocurvature fluctuations of the discrete PBH population at
formation, giving $\mathrm{S}=\delta_{\rm PBH,f}$ and isocurvature power spectrum
$\mathcal{P}_{\mathrm{S}}(k)\propto k^3$~\cite{Papanikolaou:2020qtd,Lozanov:2023aez,Lozanov:2023knf},
valid up to the fluid cutoff (ultraviolet, UV) scale $k_{\rm uv}$ set by the inverse mean PBH
separation. 
 
Below $k_{\rm uv}$ the coarse-grained fluid description of the PBH gas ceases to apply. A more restrictive cutoff is the non-linear scale $k_{\rm NL}$, where $\delta_{\rm PBH}\sim 1$. Since the gravitational potential remains perturbative with $|\Phi|\ll 1$, even when the PBH density contrast becomes nonlinear,  
we use $k_{\rm uv}$ to analyze the asymptotic scaling. The dependence on the choice of the UV cutoff is discussed in the Supplemental Material.

We numerically track the potential through the transition with a transfer function
$\Phi_k(\eta)=T_\Phi(k,\eta) \mathrm{S}_k$, where $\mathrm{S}_k$ is the initial isocurvature
amplitude.  
Because $\Phi$ is constant during eMD, modes that are already well inside the horizon remain relevant at evaporation. 
As a result, induced GWs are sourced deep inside the horizon, where the source is dominated by the time derivative of the potential,
$\mathcal{H}^{-1}\Phi'\sim c_s k\eta \Phi\gg\Phi$,
unlike those generated near horizon crossing ($k\eta\sim1$).
The late time amplitude is
inherited from the eMD plateau but modulated by the suppression factor,
$T_\Phi\propto\mathscr{S}_\Phi(k)$, which is obtained numerically in general and
reduces to the universal power law of Eq.~\eqref{eq:S_Phi_main} at high frequencies
(see Supplemental Material for details).

In Fig.~\ref{fig:OmegaGW} we display the resulting spectra for representative
LN and PL mass functions considering several evaporation indices
$\alpha$, together with the sensitivities of proposed experiments. See Supplemental Material for computational details. All cases share the
same universal high frequency suppression $\Omega_{\rm GW}\propto k^{-1/3}$ for
Hawking evaporation, which contrasts with the steeply rising $k^{11/3}$ of a monochromatic
population. This confirms that the suppression reflects the evaporation dynamics
rather than any particular formation.

The induced GWs contribute to the radiation energy density, or to effective relativistic degrees of freedom $\Delta N_{\rm eff}$~\cite{Carr:2020gox}, hence the large signal predicted in the monochromatic limit can lead to strong constraints on evaporating PBH scenarios. 
However, the universal endpoint suppression found here reduces the high frequency contribution to the integrated GW density and can substantially relax these bounds.
This generalizes the physical interpretation of Ref.~\cite{Gouttenoire:2026pem,Gouttenoire:2026pmp} beyond just the critical collapse example.

{\it{Conclusions.}}--
We have shown that finite width populations of evaporating black holes exhibit a universal endpoint evolution controlled by the evaporation law. The key physical effect is the simultaneous mass loss of individual black holes and depletion of the surviving population, which is absent in the monochromatic mass function idealization. As a result, the endpoint dynamics produces a universal asymptotic scaling behavior, insensitive to the details of the initial mass distribution.
We emphasize that this universal behavior appears for any non-monochromatic initial mass function, no matter how narrow the mass spectrum is.
In PBH-dominated early Universe scenarios, this universal behavior suppresses the gravitational potential during the transition from matter to radiation eras and imprints a strong  
modification on the scalar-induced GW spectrum. 

More broadly, our results identify induced GWs as a probe of black hole evaporation itself. 
The high frequency asymptotic spectrum carries information about the underlying mass loss law, opening a new connection between cosmological GW backgrounds, early Universe PBH populations  and black hole microphysics. 
The same reasoning may also apply to other finite  duration disappearing relic object populations, suggesting a broader class of early Universe systems in which gravitational waves can reveal the dynamics of object disappearance.

~\newline

\textit{Acknowledgments.---}
This work was supported by the World Premier International Research Center Initiative (WPI), MEXT, Japan, in part by JSPS KAKENHI Grant No.~JP24K00624, by the 111 Project (No.~B23042) and by the CSC Innovation Talent Funds.

The authors thank Yi-Fu Cai, Chao Chen, Qianhang Ding, Jinn-Ouk Gong, Qi Guo, Kazunori Kohri, Shao-Jiang Wang, and Xinpeng Wang for helpful discussions and comments. Part of this work benefited from discussions at the APCTP workshop New Perspectives on Cosmology.
\bibliography{references}

@article{Domenech:2021wkk,
    author = "Dom\`enech, Guillem and Takhistov, Volodymyr and Sasaki, Misao",
    title = "{Exploring evaporating primordial black holes with gravitational waves}",
    eprint = "2105.06816",
    archivePrefix = "arXiv",
    primaryClass = "astro-ph.CO",
    reportNumber = "IPMU21-0028, YITP-21-44",
    doi = "10.1016/j.physletb.2021.136722",
    journal = "Phys. Lett. B",
    volume = "823",
    pages = "136722",
    year = "2021"
}

@article{Ananda:2006af,
    author = "Ananda, Kishore N. and Clarkson, Chris and Wands, David",
    title = "{The Cosmological gravitational wave background from primordial density perturbations}",
    eprint = "gr-qc/0612013",
    archivePrefix = "arXiv",
    doi = "10.1103/PhysRevD.75.123518",
    journal = "Phys. Rev. D",
    volume = "75",
    pages = "123518",
    year = "2007"
}

@article{Domenech:2021ztg,
    author = "Dom{\`e}nech, Guillem",
    title = "{Scalar Induced Gravitational Waves Review}",
    eprint = "2109.01398",
    archivePrefix = "arXiv",
    primaryClass = "gr-qc",
    doi = "10.3390/universe7110398",
    journal = "Universe",
    volume = "7",
    number = "11",
    pages = "398",
    year = "2021"
}

@article{Hawking:1975vcx,
    author = "Hawking, S. W.",
    editor = "Gibbons, G. W. and Hawking, S. W.",
    title = "{Particle Creation by Black Holes}",
    doi = "10.1007/BF02345020",
    journal = "Commun. Math. Phys.",
    volume = "43",
    pages = "199--220",
    year = "1975",
    note = "[Erratum: Commun.Math.Phys. 46, 206 (1976)]"
}

@article{Carr:1975qj,
    author = "Carr, Bernard J.",
    title = "{The Primordial black hole mass spectrum}",
    doi = "10.1086/153853",
    journal = "Astrophys. J.",
    volume = "201",
    pages = "1--19",
    year = "1975"
}

@article{Khlopov:1980mg,
    author = "Khlopov, M. Yu. and Polnarev, A. G.",
    title = "{PRIMORDIAL BLACK HOLES AS A COSMOLOGICAL TEST OF GRAND UNIFICATION}",
    doi = "10.1016/0370-2693(80)90624-3",
    journal = "Phys. Lett. B",
    volume = "97",
    pages = "383--387",
    year = "1980"
}

@article{Hawking:1971ei,
    author = "Hawking, Stephen",
    title = "{Gravitationally collapsed objects of very low mass}",
    doi = "10.1093/mnras/152.1.75",
    journal = "Mon. Not. Roy. Astron. Soc.",
    volume = "152",
    pages = "75",
    year = "1971"
}

@article{Lozanov:2023rcd,
    author = "Lozanov, Kaloian D. and Pi, Shi and Sasaki, Misao and Takhistov, Volodymyr and Wang, Ao",
    title = "{Axion-like universal gravitational wave interpretation of pulsar timing array data}",
    eprint = "2310.03594",
    archivePrefix = "arXiv",
    primaryClass = "astro-ph.CO",
    reportNumber = "IPMU23-0036, YITP-23-124, KEK-QUP-2023-0025, KEK-TH-2560, KEK-Cosmo-0328, KEK-TH-2560,
  KEK-Cosmo-0328",
    doi = "10.1088/1361-6382/ad9e67",
    journal = "Class. Quant. Grav.",
    volume = "42",
    number = "3",
    pages = "035007",
    year = "2025"
}

@article{Lozanov:2023knf,
    author = "Lozanov, Kaloian D. and Sasaki, Misao and Takhistov, Volodymyr",
    title = "{Universal gravitational waves from interacting and clustered solitons}",
    eprint = "2309.14193",
    archivePrefix = "arXiv",
    primaryClass = "astro-ph.CO",
    reportNumber = "IPMU23-0033, YITP-23-115, KEK-QUP-2023-0022, KEK-TH-2555,
  KEK-Cosmo-0325",
    doi = "10.1016/j.physletb.2023.138392",
    journal = "Phys. Lett. B",
    volume = "848",
    pages = "138392",
    year = "2024"
}

@article{Lozanov:2023aez,
    author = "Lozanov, Kaloian D. and Sasaki, Misao and Takhistov, Volodymyr",
    title = "{Universal gravitational wave signatures of cosmological solitons}",
    eprint = "2304.06709",
    archivePrefix = "arXiv",
    primaryClass = "astro-ph.CO",
    reportNumber = "IPMU23-0008, KEK-QUP-2023-0006, KEK-TH-2514, KEK-Cosmo-0308,
  YITP-23-44",
    doi = "10.1088/1475-7516/2025/01/094",
    journal = "JCAP",
    volume = "01",
    pages = "094",
    year = "2025"
}

@article{Lozanov:2022yoy,
    author = "Lozanov, Kaloian D. and Takhistov, Volodymyr",
    title = "{Enhanced Gravitational Waves from Inflaton Oscillons}",
    eprint = "2204.07152",
    archivePrefix = "arXiv",
    primaryClass = "astro-ph.CO",
    reportNumber = "IPMU22-0016, KEK-QUP-2023-0008, KEK-TH-2518, KEK-Cosmo-0311",
    doi = "10.1103/PhysRevLett.130.181002",
    journal = "Phys. Rev. Lett.",
    volume = "130",
    number = "18",
    pages = "181002",
    year = "2023"
}

@article{Niemeyer:1999ak,
    author = "Niemeyer, Jens C. and Jedamzik, K.",
    title = "{Dynamics of primordial black hole formation}",
    eprint = "astro-ph/9901292",
    archivePrefix = "arXiv",
    doi = "10.1103/PhysRevD.59.124013",
    journal = "Phys. Rev. D",
    volume = "59",
    pages = "124013",
    year = "1999"
}

@article{Evans:1994pj,
    author = "Evans, Charles R. and Coleman, Jason S.",
    title = "{Observation of critical phenomena and selfsimilarity in the gravitational collapse of radiation fluid}",
    eprint = "gr-qc/9402041",
    archivePrefix = "arXiv",
    reportNumber = "TAR-039-UNC",
    doi = "10.1103/PhysRevLett.72.1782",
    journal = "Phys. Rev. Lett.",
    volume = "72",
    pages = "1782--1785",
    year = "1994"
}

@article{Choptuik:1992jv,
    author = "Choptuik, Matthew W.",
    title = "{Universality and scaling in gravitational collapse of a massless scalar field}",
    reportNumber = "FPRINT-92-33",
    doi = "10.1103/PhysRevLett.70.9",
    journal = "Phys. Rev. Lett.",
    volume = "70",
    pages = "9--12",
    year = "1993"
}

@article{Dolgov:1992pu,
    author = "Dolgov, Alexandre and Silk, Joseph",
    title = "{Baryon isocurvature fluctuations at small scales and baryonic dark matter}",
    reportNumber = "CFPA-TH-92-04",
    doi = "10.1103/PhysRevD.47.4244",
    journal = "Phys. Rev. D",
    volume = "47",
    pages = "4244--4255",
    year = "1993"
}

@article{Kannike:2017bxn,
    author = {Kannike, Kristjan and Marzola, Luca and Raidal, Martti and Veerm{\"a}e, Hardi},
    title = "{Single Field Double Inflation and Primordial Black Holes}",
    eprint = "1705.06225",
    archivePrefix = "arXiv",
    primaryClass = "astro-ph.CO",
    doi = "10.1088/1475-7516/2017/09/020",
    journal = "JCAP",
    volume = "09",
    pages = "020",
    year = "2017"
}

@article{Carr:1974nx,
    author = "Carr, Bernard J. and Hawking, S. W.",
    title = "{Black holes in the early Universe}",
    doi = "10.1093/mnras/168.2.399",
    journal = "Mon. Not. Roy. Astron. Soc.",
    volume = "168",
    pages = "399--415",
    year = "1974"
}

@article{Zeldovich:1967lct,
    author = "Zel'dovich, Ya. B. and Novikov, I. D.",
    title = "{The Hypothesis of Cores Retarded during Expansion and the Hot Cosmological Model}",
    journal = "Sov. Astron.",
    volume = "10",
    pages = "602",
    year = "1967"
}

@article{Hawking:1974rv,
    author = "Hawking, S. W.",
    title = "{Black hole explosions}",
    doi = "10.1038/248030a0",
    journal = "Nature",
    volume = "248",
    pages = "30--31",
    year = "1974"
}

@article{Baumann:2007zm,
    author = "Baumann, Daniel and Steinhardt, Paul J. and Takahashi, Keitaro and Ichiki, Kiyotomo",
    title = "{Gravitational Wave Spectrum Induced by Primordial Scalar Perturbations}",
    eprint = "hep-th/0703290",
    archivePrefix = "arXiv",
    doi = "10.1103/PhysRevD.76.084019",
    journal = "Phys. Rev. D",
    volume = "76",
    pages = "084019",
    year = "2007"
}

@article{He:2024luf,
    author = "He, Xin-Chen and Cai, Yi-Fu and Ma, Xiao-Han and Papanikolaou, Theodoros and Saridakis, Emmanuel N. and Sasaki, Misao",
    title = "{Gravitational waves from primordial black hole isocurvature: the effect of non-Gaussianities}",
    eprint = "2409.11333",
    archivePrefix = "arXiv",
    primaryClass = "astro-ph.CO",
    reportNumber = "YITP-24-93",
    doi = "10.1088/1475-7516/2024/12/039",
    journal = "JCAP",
    volume = "12",
    pages = "039",
    year = "2024"
}

@article{Inomata:2020lmk,
    author = "Inomata, Keisuke and Kawasaki, Masahiro and Mukaida, Kyohei and Terada, Takahiro and Yanagida, Tsutomu T.",
    title = "{Gravitational Wave Production right after a Primordial Black Hole Evaporation}",
    eprint = "2003.10455",
    archivePrefix = "arXiv",
    primaryClass = "astro-ph.CO",
    reportNumber = "IPMU 20-0029, DESY 20-042, DESY-20-042, CTPU-PTC-20-05",
    doi = "10.1103/PhysRevD.101.123533",
    journal = "Phys. Rev. D",
    volume = "101",
    number = "12",
    pages = "123533",
    year = "2020"
}

@article{Carr:2020gox,
    author = "Carr, Bernard and Kohri, Kazunori and Sendouda, Yuuiti and Yokoyama, Jun'ichi",
    title = "{Constraints on primordial black holes}",
    eprint = "2002.12778",
    archivePrefix = "arXiv",
    primaryClass = "astro-ph.CO",
    reportNumber = "RESCEU-03/20; KEK-Cosmo-249; KEK-TH-2199; IPMU20-0024",
    doi = "10.1088/1361-6633/ac1e31",
    journal = "Rept. Prog. Phys.",
    volume = "84",
    number = "11",
    pages = "116902",
    year = "2021"
}

@article{Kohri:2018awv,
    author = "Kohri, Kazunori and Terada, Takahiro",
    title = "{Semianalytic calculation of gravitational wave spectrum nonlinearly induced from primordial curvature perturbations}",
    eprint = "1804.08577",
    archivePrefix = "arXiv",
    primaryClass = "gr-qc",
    reportNumber = "KEK-TH-2046, KEK-COSMO-223",
    doi = "10.1103/PhysRevD.97.123532",
    journal = "Phys. Rev. D",
    volume = "97",
    number = "12",
    pages = "123532",
    year = "2018"
}

@article{Domenech:2020ssp,
    author = "Dom\`enech, Guillem and Lin, Chunshan and Sasaki, Misao",
    title = "{Gravitational wave constraints on the primordial black hole dominated early universe}",
    eprint = "2012.08151",
    archivePrefix = "arXiv",
    primaryClass = "gr-qc",
    reportNumber = "YITP-20-156",
    doi = "10.1088/1475-7516/2021/11/E01",
    journal = "JCAP",
    volume = "04",
    pages = "062",
    year = "2021",
    note = "[Erratum: JCAP 11, E01 (2021)]"
}

@article{Papanikolaou:2020qtd,
    author = "Papanikolaou, Theodoros and Vennin, Vincent and Langlois, David",
    title = "{Gravitational waves from a universe filled with primordial black holes}",
    eprint = "2010.11573",
    archivePrefix = "arXiv",
    primaryClass = "astro-ph.CO",
    doi = "10.1088/1475-7516/2021/03/053",
    journal = "JCAP",
    volume = "03",
    pages = "053",
    year = "2021"
}

@article{Domenech:2024wao,
    author = {Dom\`enech, Guillem and Tr\"ankle, Jan},
    title = "{From formation to evaporation: Induced gravitational wave probes of the primordial black hole reheating scenario}",
    eprint = "2409.12125",
    archivePrefix = "arXiv",
    primaryClass = "gr-qc",
    month = "9",
    year = "2024"
}

@article{Papanikolaou:2022chm,
    author = "Papanikolaou, Theodoros",
    title = "{Gravitational waves induced from primordial black hole fluctuations: the~effect of an extended mass function}",
    eprint = "2207.11041",
    archivePrefix = "arXiv",
    primaryClass = "astro-ph.CO",
    doi = "10.1088/1475-7516/2022/10/089",
    journal = "JCAP",
    volume = "10",
    pages = "089",
    year = "2022"
}

@article{Carr:2016hva,
    author = "Carr, B. J. and Kohri, Kazunori and Sendouda, Yuuiti and Yokoyama, Jun'ichi",
    title = "{Constraints on primordial black holes from the Galactic gamma-ray background}",
    eprint = "1604.05349",
    archivePrefix = "arXiv",
    primaryClass = "astro-ph.CO",
    reportNumber = "RESCEU-16-16, KEK-TH-1895, KEK-COSMO-193",
    doi = "10.1103/PhysRevD.94.044029",
    journal = "Phys. Rev. D",
    volume = "94",
    number = "4",
    pages = "044029",
    year = "2016"
}

@article{Gouttenoire:2026pmp,
    author = "Gouttenoire, Yann and Leister, Nicholas and Schwaller, Pedro",
    title = "{Gravitational Waves from Black Hole Reheating: The Scalar-Induced Component}",
    eprint = "2605.21474",
    archivePrefix = "arXiv",
    primaryClass = "hep-ph",
    reportNumber = "MITP-26-024",
    month = "5",
    year = "2026"
}

@article{Inomata:2019ivs,
    author = "Inomata, Keisuke and Kohri, Kazunori and Nakama, Tomohiro and Terada, Takahiro",
    title = "{Enhancement of Gravitational Waves Induced by Scalar Perturbations due to a Sudden Transition from an Early Matter Era to the Radiation Era}",
    eprint = "1904.12879",
    archivePrefix = "arXiv",
    primaryClass = "astro-ph.CO",
    reportNumber = "IPMU 19-0067, KEK-TH-2122, KEK-Cosmo-237",
    doi = "10.1103/PhysRevD.108.049901",
    journal = "Phys. Rev. D",
    volume = "100",
    pages = "043532",
    year = "2019",
    note = "[Erratum: Phys.Rev.D 108, 049901 (2023)]"
}

@article{Gouttenoire:2026pem,
    author = "Gouttenoire, Yann and Leister, Nicholas and Schwaller, Pedro",
    title = "{Opening the Window of Ultra-Light PBHs by Exorcising the Poltergeist}",
    eprint = "2605.21477",
    archivePrefix = "arXiv",
    primaryClass = "hep-ph",
    reportNumber = "MITP-26-023",
    month = "5",
    year = "2026"
}

@article{Kusenko:1997si,
    author = "Kusenko, Alexander and Shaposhnikov, Mikhail E.",
    title = "{Supersymmetric Q balls as dark matter}",
    eprint = "hep-ph/9709492",
    archivePrefix = "arXiv",
    reportNumber = "CERN-TH-97-259",
    doi = "10.1016/S0370-2693(97)01375-0",
    journal = "Phys. Lett. B",
    volume = "418",
    pages = "46--54",
    year = "1998"
}

@article{Maggiore:2019uih,
    author = "Maggiore, Michele and others",
    title = "{Science Case for the Einstein Telescope}",
    eprint = "1912.02622",
    archivePrefix = "arXiv",
    primaryClass = "astro-ph.CO",
    doi = "10.1088/1475-7516/2020/03/050",
    journal = "JCAP",
    volume = "03",
    pages = "050",
    year = "2020"
}

@article{Harry:2006fi,
    author = "Harry, G. M. and Fritschel, P. and Shaddock, D. A. and Folkner, W. and Phinney, E. S.",
    title = "{Laser interferometry for the big bang observer}",
    doi = "10.1088/0264-9381/23/15/008",
    journal = "Class. Quant. Grav.",
    volume = "23",
    pages = "4887--4894",
    year = "2006",
    note = "[Erratum: Class.Quant.Grav. 23, 7361 (2006)]"
}

@article{Karnesis:2022vdp,
    author = "Karnesis, Nikolaos and others",
    title = "{The Laser Interferometer Space Antenna mission in Greece White Paper}",
    eprint = "2209.04358",
    archivePrefix = "arXiv",
    primaryClass = "gr-qc",
    month = "9",
    year = "2022"
}

@article{LISA:2017pwj,
    author = "Amaro-Seoane, Pau and others",
    collaboration = "LISA",
    title = "{Laser Interferometer Space Antenna}",
    eprint = "1702.00786",
    archivePrefix = "arXiv",
    primaryClass = "astro-ph.IM",
    month = "2",
    year = "2017"
}

@article{LIGOScientific:2025bgj,
    author = "Abac, A. G. and others",
    collaboration = "LIGO Scientific, VIRGO, KAGRA",
    title = "{Upper Limits on the Isotropic Gravitational-Wave Background from the first part of LIGO, Virgo, and KAGRA's fourth Observing Run}",
    eprint = "2508.20721",
    archivePrefix = "arXiv",
    primaryClass = "gr-qc",
    reportNumber = "LIGO-P2500349",
    month = "8",
    year = "2025"
}

@article{Papanikolaou:2024kjb,
    author = "Papanikolaou, Theodoros and He, Xin-Chen and Ma, Xiao-Han and Cai, Yi-Fu and Saridakis, Emmanuel N. and Sasaki, Misao",
    title = "{New probe of non-Gaussianities with primordial black hole induced gravitational waves}",
    eprint = "2403.00660",
    archivePrefix = "arXiv",
    primaryClass = "astro-ph.CO",
    reportNumber = "YITP-24-22",
    doi = "10.1016/j.physletb.2024.138997",
    journal = "Phys. Lett. B",
    volume = "857",
    pages = "138997",
    year = "2024"
}

@article{Pearce:2025ywc,
    author = "Pearce, Matthew and Pearce, Lauren and White, Graham and Bal{\'a}zs, Csaba",
    title = "{Using gravitational wave signals to disentangle early matter dominated epochs}",
    eprint = "2503.03101",
    archivePrefix = "arXiv",
    primaryClass = "astro-ph.CO",
    doi = "10.1088/1475-7516/2025/11/004",
    journal = "JCAP",
    volume = "11",
    pages = "004",
    year = "2025"
}

@article{Pearce:2023kxp,
    author = "Pearce, Matthew and Pearce, Lauren and White, Graham and Balazs, Csaba",
    title = "{Gravitational wave signals from early matter domination: interpolating between fast and slow transitions}",
    eprint = "2311.12340",
    archivePrefix = "arXiv",
    primaryClass = "astro-ph.CO",
    doi = "10.1088/1475-7516/2024/06/021",
    journal = "JCAP",
    volume = "06",
    pages = "021",
    year = "2024"
}

@article{Cai:2021fgm,
    author = "Cai, Yi-Fu and Chen, Chao and Ding, Qianhang and Wang, Yi",
    title = "{Cosmological standard timers from unstable primordial relics}",
    eprint = "2112.10422",
    archivePrefix = "arXiv",
    primaryClass = "astro-ph.CO",
    doi = "10.1140/epjc/s10052-023-12046-0",
    journal = "Eur. Phys. J. C",
    volume = "83",
    number = "10",
    pages = "913",
    year = "2023"
}

@article{Chiba:2009zu,
    author = "Chiba, Takeshi and Kamada, Kohei and Yamaguchi, Masahide",
    title = "{Gravitational Waves from Q-ball Formation}",
    eprint = "0912.3585",
    archivePrefix = "arXiv",
    primaryClass = "astro-ph.CO",
    reportNumber = "RESCEU-30-09",
    doi = "10.1103/PhysRevD.81.083503",
    journal = "Phys. Rev. D",
    volume = "81",
    pages = "083503",
    year = "2010"
}

\clearpage
\newpage  
\appendix
\onecolumngrid
\renewcommand{\theequation}{S.\arabic{equation}}
\setcounter{equation}{0}

\centerline{\large {Supplemental Material for}}
\medskip

{\centerline{\large \bf{Universal Suppression of Gravitational Waves from Black Hole Evaporation Dynamics}}}
\medskip
{\centerline{Xin-Chen He, Xiao-Han Ma, Misao Sasaki, Volodymyr Takhistov}}
\bigskip
\bigskip
  
\vspace{1em} 
\counterwithin{figure}{section}
\counterwithin{table}{section}

In this Supplemental Material we provide additional details on calculation of the PBH mass function evolution, universal scaling from evaporation, PBH number density and energy density scaling, suppression factor of gravitational potential and induced gravitational waves.
 
\section{A. Evolution of Extended Mass Functions}

To derive the evolution equation for PBH population $f(M,t)$, it is convenient to consider evaporation of PBH population as a flow in mass space. The function $f(M,t)$ is defined
so that $f(M,t) dM$ is the comoving number density of objects with mass in
$[M,M+dM]$ at time $t$. Then, the comoving number density is
\begin{equation}
n_{\rm PBH,c}(t)=\int_0^\infty f(M,t) dM~.
\end{equation}
Black hole evaporation results in mass loss and evolves each object
continuously through mass space towards lower $M$, without creating or
destroying objects before they reach $M\to 0$.

For a Schwarzschild PBH, Hawking emission gives the mass loss rate of
\begin{equation}
    \dot M(M)=-\frac{A M_{\rm pl}^4}{M^2}=-\Gamma_1(M) M ,
    \qquad
    A=\frac{3.8\pi g_H(T_{\rm PBH})}{480},
    \label{eq:hawking_rate}
\end{equation}
where $\Gamma_1(M)$ is the evaporation rate of a single PBH and $g_H$ counts
the radiated degrees of freedom. Generalizing this, we can consider
\begin{equation}
    \dot M(M)=-\kappa M^{-\alpha}, 
    \label{eq:generalized_rate}
\end{equation}
with $\kappa>0$ and $\alpha>-1$,
which reduces to Hawking evaporation for $\alpha=2$, $\kappa=A M_{\rm pl}^4$ and other values describe modification of evaporation law. In all cases we consider $\dot M<0$.

Let us consider a  fixed  mass interval $[M_1,M_2]$, then the number density of objects it
contains is
\begin{equation}
    n_{[M_1,M_2]}(t)=\int_{M_1}^{M_2} f(M,t) dM .
\end{equation}
Its time derivative is given by 
\begin{equation}
    \frac{dn_{[M_1,M_2]}}{dt}
    =\int_{M_1}^{M_2}\frac{\partial f(M,t)}{\partial t} dM .
    \label{eq:lhs_fixed}
\end{equation}

The PBH number inside a fixed mass interval can change only because objects flow across
its boundaries during evolution. The mass space velocity $\dot M(M)$ gives
the number density crossing a given mass value per unit  comoving  volume and time is
the flux $\dot M(M) f(M,t)$. Balancing the flux through the 
fixed interval boundaries  one has
\begin{equation}
    \frac{dn_{[M_1,M_2]}}{dt}
     =\dot M(M_1)f(M_1,t)-\dot M(M_2)f(M_2,t) = -\int_{M_1}^{M_2}\frac{\partial}{\partial M} ( \dot M(M)f(M,t)) dM 
    \label{eq:flux_balance}
\end{equation}
Since $\dot M<0$, the term at $M_1$ is an outflow through the lower boundary
and the term at $M_2$ an inflow from higher masses. Equating
\eqref{eq:lhs_fixed} and \eqref{eq:flux_balance} gives
\begin{equation}
    \int_{M_1}^{M_2}
    \left[\frac{\partial f}{\partial t}
    +\frac{\partial}{\partial M}\big(\dot M(M)f\big)\right]dM=0 .
\end{equation}
As $M_1,M_2$ are arbitrary, the integrand vanishes identically, yielding the local 
continuity equation in mass space 
\begin{equation}
    \frac{\partial f(M,t)}{\partial t}
    +\frac{\partial}{\partial M}\big[\dot M(M) f(M,t)\big]=0 ,
    \label{eq:continuity_SM}
\end{equation}
which corresponds to fluid continuity equation considering $\dot M(M)$ as the flow velocity.
In general the evaporation rate may also depend on additional  
parameters such as black hole spin or charge. Then, Eq.~\eqref{eq:continuity_SM} would be
modified to account for the full parameter space. Here, we restrict our analysis to an effective
one dimensional description governed by $\dot M=\dot M(M)$.

Eq.~\eqref{eq:continuity_SM} can be solved by the method of characteristics. The
characteristic trajectories of individual objects in mass space are
\begin{equation}
    \frac{dM}{dt}=\dot M(M),
    \label{eq:characteristic}
\end{equation}
labeled by their initial mass $M_0= M(t{=}0)$, which can be used to consider  
the initial mass of an object that has mass $M$ at time $t$.
Since characteristic trajectories do not cross, the objects in the interval $[M_0,M_0+dM_0]$ at $t=0$
are those in the interval $[M,M+dM]$ at time $t$, and their number is conserved. This gives
\begin{equation}
    f(M,t) dM=f_0(M_0) dM_0 ,
    \qquad f_0(M)= f(M,0)~.
    \label{eq:number_conservation}
\end{equation}

Thus, $f(M,t)=f_0(M_0)\left(dM_0/dM\right)|_t$. To determine the Jacobian, we can consider   neighboring characteristic trajectories. The separation between them is proportional to the local mass space velocity. Hence
\begin{equation}
    \frac{dM}{\dot M(M)}
    =
    \frac{dM_0}{\dot M(M_0)} ,
\end{equation}
which gives
\begin{equation}
    \left(\frac{dM_0}{dM}\right)\Bigg|_t
    =
    \frac{\dot M(M_0)}{\dot M(M)} .
\end{equation}
Therefore, we obtain general result
\begin{equation}
    f(M,t)
    =
    f_0 \left(M_0(M,t)\right)
    \frac{\dot M \left(M_0(M,t)\right)}
         {\dot M(M)} .
    \label{eq:solution_of_f}
\end{equation}

\section{B. Universal Scaling from Evaporation}
\label{sec:universal_scaling}

We now consider the general power law evaporation of Eq.~\eqref{eq:generalized_rate}, with Hawking evaporation being its special $\alpha = 2$ case, and
show that the low mass tail of $f(M,t)$ approaches a universal scaling set by the
evaporation law alone. Integrating Eq.~\eqref{eq:characteristic}
with $\dot M=-\kappa M^{-\alpha}$ gives
\begin{equation}
    M_0(M,t)=\big[M^{\alpha+1}+(\alpha+1)\kappa t\big]^{\frac{1}{\alpha+1}}.
    \label{eq:M0_powerlaw}
\end{equation}
The general solution Eq.~\eqref{eq:solution_of_f} becomes
\begin{equation}
    f(M,t)=\left(\frac{M}{M_0(M,t)}\right)^{ \alpha} f_0\big(M_0(M,t)\big).
    \label{eq:f_powerlaw}
\end{equation}
For any $t>0$, in the low mass regime $M^{\alpha+1}\ll(\alpha+1)\kappa t$ the initial
mass $M_0$ becomes   independent of $M$, such that
$M_0(M,t)\to[(\alpha+1)\kappa t]^{1/(\alpha+1)}$. Thus, Eq.~\eqref{eq:f_powerlaw}
reduces to the universal low mass scaling
\begin{equation}
    f(M,t)\;\simeq\;
    \frac{f_0\big([(\alpha+1)\kappa t]^{1/(\alpha+1)}\big)}
         {[(\alpha+1)\kappa t]^{\alpha/(\alpha+1)}} M^{\alpha}
    \;\propto\; M^{\alpha}.
    \label{eq:universal_scaling}
\end{equation}
This holds without taking the strict limit $M\to0$, and no assumption was made
about $f_0$. The distribution tail is fixed entirely by the evaporation law, since the
rapidly growing rate near $M\to0$ depletes the low mass population. The
argument here requires $\alpha>-1$, which accounts also for standard Hawking evaporation and which we assume throughout.

\section{C. PBH Number and Energy Density Scaling}
\label{sec:late_time_scaling}

We compute the asymptotic scaling of the comoving PBH number and energy densities near complete evaporation. Assuming that the initial mass function has an effective maximum mass $M_{\rm max}$, the final evaporation time is
\begin{equation}
    t_{\rm evap}
    =
    \frac{M_{\rm max}^{\alpha+1}}{(\alpha+1)\kappa}.
\end{equation}
The following scaling assumes that the initial mass function is finite, non-vanishing  and sufficiently slowly varying near this effective endpoint.

At time $t<t_{\rm evap}$, the largest surviving PBH mass is the evolved mass of an object whose initial mass was $M_{\rm max}$. Thus
\begin{equation}
    M_{\rm top}(t)
    =
    \left[
    M_{\rm max}^{\alpha+1}
    -
    (\alpha+1)\kappa t
    \right]^{\frac{1}{\alpha+1}}
    =
    \left[
    (\alpha+1)\kappa (t_{\rm evap}-t)
    \right]^{\frac{1}{\alpha+1}} .
    \label{eq:M_top}
\end{equation}
Near $t_{\rm evap}$  the integration region $0<M<M_{\rm top}(t)$ shrinks to low final PBH masses. The corresponding initial masses approach the upper edge of the initial distribution with $M_0(M,t)\simeq M_{\rm max}$. Considering Eq.~\eqref{eq:f_powerlaw} we have
\begin{equation}
    f(M,t)
    \simeq \Big(
    \frac{ M }{M_{\rm max} } \Big)^{\alpha}    f_0(M_{\rm max}) .
    \label{eq:f_endpoint}
\end{equation}

The comoving number density of surviving PBHs is then
\begin{equation}
    n_{\rm PBH,c}(t)
    =
    \int_0^{M_{\rm top}(t)} f(M,t) dM
    \simeq
    \frac{\left[M_{\rm top}(t)\right]^{\alpha+1}}{(\alpha+1)M_{\rm max}^{\alpha}}
     f_0(M_{\rm max}).
\end{equation}
Using Eq.~\eqref{eq:M_top}, this becomes
\begin{equation}
    n_{\rm PBH,c}(t)
    \simeq
    \frac{\kappa f_0(M_{\rm max})}{M_{\rm max}^{\alpha}}
    (t_{\rm evap}-t).
    \label{eq:n_endpoint}
\end{equation}
Thus, the comoving number of surviving PBHs decreases linearly as $\propto (t_{\rm evap} - t)$ near the endpoint. This linear depletion is independent of $\alpha$ and is absent in the monochromatic limit, where the number density remains constant until the common evaporation time.

Similarly, the comoving PBH energy density is
\begin{equation}
    \rho_{\rm PBH,c}(t)
    =
    \int_0^{M_{\rm top}(t)} M f(M,t) dM
    \simeq
    \frac{ \left[M_{\rm top}(t)\right]^{\alpha+2}}{(\alpha+2)M_{\rm max}^{\alpha}}
    f_0(M_{\rm max}).
\end{equation}
Therefore,
\begin{equation}
    \rho_{\rm PBH,c}(t)
    \propto
    (t_{\rm evap}-t)^{\frac{\alpha+2}{\alpha+1}}
    =
    (t_{\rm evap}-t)^{1+\frac{1}{\alpha+1}} .
    \label{eq:rhoPBH_scaling}
\end{equation}
Equivalently, this scaling follows from $\rho_{\rm PBH,c}\sim n_{\rm PBH,c}M$, with $n_{\rm PBH,c}\propto(t_{\rm evap}-t)$ and $M\propto(t_{\rm evap}-t)^{1/(\alpha+1)}$. For standard Hawking evaporation with $\alpha=2$, this gives $\rho_{\rm PBH,c}\propto(t_{\rm evap}-t)^{4/3}$.

\section{D. Suppression Factor of Gravitational Potential}
\label{sec:suppression_factor}
 
We discuss further considerations  underlying the high $k$ suppression factor in the main text. For sub-horizon modes during the PBH-dominated stage, the Poisson equation gives
\begin{equation}
k^2\Phi
\sim
\dfrac{a^2\rho_{\rm PBH}\delta_{\rm PBH}}{M_{\rm Pl}^2}.
\label{eq}
\end{equation}
Prior to significant evaporation PBHs behave as pressureless matter. In the linear regime $\delta_{\rm PBH}\propto a$ and with $\rho_{\rm PBH}=a^{-3}\rho_{\rm PBH,c}$ one obtains $\Phi\propto \rho_{\rm PBH,c}$ up to factors that vary slowly during the eMD plateau. In the absence of an additional rapid nonlinear process, the leading endpoint suppression of $\Phi$ is controlled by $\rho_{\rm PBH,c}$.

The collective evaporation rate is defined as
\begin{equation}
\Gamma(t)
=
-
\frac{\dot{\rho}_{\rm PBH,c}}{\rho_{\rm PBH,c}} .
\label{eq}
\end{equation}
Using Eq.~\eqref{eq:rhoPBH_scaling}, near complete evaporation one has
\begin{equation}
\Gamma(t)
=
\frac{\alpha+2}{\alpha+1}
(t_{\rm evap}-t)^{-1}.
\label{eq:gamma}
\end{equation}

The potential follows the PBH density perturbation until the PBH component changes appreciably within  oscillation time of the mode. This defines the decoupling time through
\begin{equation}
\Gamma(t_{\rm dec})
\simeq
\frac{k}{a(t_{\rm dec})}.
\label{eq}
\end{equation}
For the high $k$ modes of interest, $t_{\rm dec}$ lies very close to $t_{\rm evap}$, so $a(t_{\rm dec})\simeq a_{\rm evap}$. Combining this with Eq.~\eqref{eq:gamma} gives
\begin{equation} 
(t_{\rm evap}-t_{\rm dec}(k))
\propto
k^{-1}.
\label{eq:timescale}
\end{equation}

The suppression factor is defined relative to the eMD plateau value of the potential. Since $\Phi$ tracks the PBH density perturbation until $t_{\rm dec}$, its high $k$ scaling is fixed by the comoving PBH energy density at decoupling 
\begin{equation}
\mathscr{S}_\Phi(k)
\propto
\rho_{\rm PBH,c}(t_{\rm dec}).
\label{eq}
\end{equation}
Using Eq.~\eqref{eq:rhoPBH_scaling} together with Eq.~\eqref{eq:timescale}, one obtains
\begin{equation}
\mathscr{S}_\Phi(k)
\propto
k^{-\frac{\alpha+2}{\alpha+1}}
=
k^{-1-\frac{1}{\alpha+1}} .
\label{eq:suppscale}
\end{equation}
For Hawking evaporation, $\alpha=2$, this reduces to $\mathscr{S}_\Phi(k)\propto k^{-4/3}$. The additional factor of $k^{-1}$ relative to the monochromatic case originates from the linear depletion of the surviving PBH number density near the endpoint.

\section{E. Induced Gravitational Waves}
\subsubsection{E1. Evolution of gravitational potential}

For a finite width PBH mass function the population evaporates over an extended
period of time. The evolution of the scale factor and the Hubble parameter can be well approximated by their forms during the RD era, namely, $a(\eta)\propto \eta$ and ${\cal H}(\eta)\simeq1/\eta$. Consequently, the gravitational potential follows RD era evolution
\begin{equation}
    \Phi''+\frac{4}{\eta}\Phi'+c_s^2k^2\Phi=0~,
\end{equation}
where $c_s=1/\sqrt{3}$ is cosmological fluid sound speed. We consider initial conditions $\Phi_k(\eta_{\rm evap})=\Phi_{{\rm evap}}(k),~\Phi'(\eta_{\rm evap})=\Phi_{{\rm evap}}'(k)$. With $x= k\eta$ and taking the sub-horizon limit $x_{\rm evap}= k\eta_{\rm evap}\gg1$,  we obtain
\begin{eqnarray}
    \Phi_k(\eta)\simeq\left(\frac{x_{\rm evap}}{x}\right)^2\left(\Phi_{\rm evap}(k)\cos\big(c_s(x-x_{\rm evap})\big)+\frac{\Phi'_{\rm evap}(k)}{c_s k}\sin\big(c_s(x-x_{\rm evap})\big)\right)~.
\end{eqnarray}
Its derivative with respect to $x$  at leading order is
\begin{eqnarray}
    \frac{d}{dx}\Phi_k(\eta)\simeq\left(\frac{x_{\rm evap}}{x}\right)^2 c_s \left(-\Phi_{\rm evap}(k)\sin\big(c_s(x-x_{\rm evap})\big)+\frac{\Phi'_{\rm evap}(k)}{c_s k}\cos\big(c_s(x-x_{\rm evap})\big)\right)~.
\end{eqnarray}
Since the Universe is already radiation dominated at the complete evaporation time $\eta_{\rm evap}$, we must have $|\Phi_k'|\simeq c_s k |\Phi_k|$ for $\eta\geq\eta_{\rm evap}$ where prime denotes derivative with respect to $\eta$. This implies ${\cal H}^{-1}|\Phi'|\simeq \eta|\Phi'|\simeq c_s k\eta|\Phi|\gg|\Phi|$. Consequently, the ${\cal H}^{-1}\Phi'$ contribution dominates as the induced GW source.  

Let us consider the initial conditions for the gravitational potential at the evaporation time, $\Phi_{\rm evap}(k)$ and $\Phi_{\rm evap}'(k)$. 
$\Phi_{\rm evap}(k)$ can be parameterized as
\begin{eqnarray}\label{eq:Phi_evap}
    \Phi_{\rm evap}(k)=-\mathscr{S}_{\Phi}(k) T_{\Phi,{\rm eMD}}(k)\mathrm{S}_k\cos\phi_k~, 
\end{eqnarray}
where $\mathrm{S}_k$ denotes the initial PBH isocurvature fluctuation and $\mathscr{S}_\Phi(k)$ describes the suppression during the later PBH evaporation and eMD-RD transition.
Here, $T_{\rm eMD}(k)$ denote the transfer from the early RD era to subsequent PBH eMD  era 
\begin{eqnarray}
    T_{\Phi,\rm eMD}(k)\simeq C(w)\left(\frac{k}{k_{\rm d}}\right)^{-2}~,
\end{eqnarray}
where $k_{\rm d}$  denotes the horizon scale at the onset of PBH domination, $w$ is the equation of state parameter before PBH domination and  we adopt the same notation for $C(w)$ as Ref.~\cite{Domenech:2024wao}, with $C(1/3)\simeq1.11$. 

The suppression factor $\mathscr{S}_\Phi(k)$ is the ratio of the potential after evaporation to its eMD plateau value. In the high frequency regime, its scaling is fixed by the evaporation endpoint result as derived in Eq.~\eqref{eq:suppscale}. We therefore parameterize \begin{equation} \mathscr{S}_\Phi(k) = \mathcal{A}_\Phi \left(\frac{k}{k_{\rm evap}}\right)^{-b}, \label{eq:Sphi_powerlaw} \end{equation} where \begin{equation} b= \begin{cases} \dfrac{1}{\alpha+1}, & \text{monochromatic mass function}, \\[8pt] 1+\dfrac{1}{\alpha+1}, & \text{finite width mass function}. \end{cases} \label{eq:b_exponent} \end{equation} 
For Hawking evaporation with $\alpha=2$  this gives $b=1/3$ in the monochromatic limit and $b=4/3$ for finite width distributions. The exponent $b$ is controlled by the evaporation law, while the normalization ${\cal A}_\Phi$ depends on the detailed mass function, the transition history  and the definition of $k_{\rm evap}$. We therefore determine ${\cal A}_\Phi$ numerically for the spectra shown in the main text.

The phase $\phi_k$ in Eq.~\eqref{eq:Phi_evap} encodes the oscillatory phase of the mode $k$ at the onset of the post-evaporation RD era. Since the subsequent evolution is oscillatory, $\Phi_{\rm evap}$ and $\Phi'_{\rm evap}$ can be written in terms of the same amplitude with a relative $\pi/2$ phase shift up to a sign convention. We therefore parameterize
\begin{equation}
    \frac{\Phi_{\rm evap}'(k)}{c_s k}
    =
    \mathscr{S}_{\Phi}(k) 
    T_{\Phi,{\rm eMD}}(k) 
    \sin\phi_k \mathrm{S}_{k} .
    \label{eq:Phi_prime_evap}
\end{equation}
Defining the post-evaporation transfer function by
$\Phi_k(\eta>\eta_{\rm evap})=T_{\Phi,{\rm post}}(k,\eta)\mathrm{S}_{k}$, the leading sub-horizon derivative contribution becomes
\begin{equation}
    \frac{d T_{\Phi,{\rm post}}(k,x)}{d x}
    =
    \left(\frac{x_{\rm evap}}{x}\right)^2
    c_s \mathscr{S}_\Phi(k) 
    T_{\Phi,{\rm eMD}}(k)
    \sin \left[c_s(x-x_{\rm evap})+\phi_k\right] .
    \label{eq:dTpost_dx}
\end{equation}

\subsubsection{E2. Kernel}

The gravitational potential continuously sources GWs at second-order. The GW energy spectrum is ~\cite{Ananda:2006af,Baumann:2007zm,Kohri:2018awv,Domenech:2021ztg}
 \begin{equation} 
\label{eq:OmegaGW} 
\Omega_{\rm GW}(\eta,k) = \frac{2k^2}{3{\cal H}^2}\int_{0}^{\infty} 
\mathrm{d}v\int_{|1-v|}^{1+v}\mathrm{d}u   
\left[ \frac{4v^2 \!-\! (1\!+\!v^2\!-\!u^2)^2}{4uv}\right]^{2} 
\overline{I^2(u,v,k,\eta)}\mathcal{P}_{\mathrm{S}}(kv)\mathcal{P}_{\mathrm{S}}(ku) .
\end{equation}
where $\eta$ is conformal time, and the overbar denotes time averaging. The integration variables are $u=|{\bf k}-{\bf q}|/k$ and $v=q/k$. The kernel $I(u,v,k,\eta)$ is determined by the time evolution of the scalar transfer function $T(k,\eta)$ and accounts for the continuous sourcing of GWs by scalar perturbations.

We next evaluate the contribution to the scalar induced GW kernel. The time integrated kernel is \begin{equation} 
I(u,v,k,\eta) = k\int_0^\eta d\tilde{\eta}  \frac{a(\tilde{\eta})}{a(\eta)} kG_k(\eta,\tilde{\eta})  F(u,v,k,\tilde{\eta}) , \label{eq:kernel_def} 
\end{equation} 
where  during RD era
\begin{equation}   F(u,v,k,\tilde{\eta}) =   T_\Phi(vk,\tilde{\eta})T_\Phi(uk,\tilde{\eta})  + \frac{1}{2} \left[ T_\Phi(vk,\tilde{\eta}) + \frac{T_\Phi'(vk,\tilde{\eta})}{\mathcal{H}} \right] \left[ T_\Phi(uk,\tilde{\eta}) + \frac{T_\Phi'(uk,\tilde{\eta})}{\mathcal{H}} \right] .  \label{eq:source_function} \end{equation} 
Here, $T_\Phi$ is the full transfer function   $\Phi_k(\eta)=T_\Phi(k,\eta) \mathrm{S}_{k}$ and for post-evaporation contribution we use the form $T_{\Phi, {\rm post}}$.
The retarded Green's function in RD is \begin{equation} kG_k(\eta,\tilde{\eta}) = \sin \left[k(\eta-\tilde{\eta})\right] . \label{eq:RD_green} \end{equation}

For the high frequency modes of interest, the dominant contribution comes from the post-evaporation oscillatory regime. In this regime $k\eta_{\rm evap}\gg1$  and the scalar source is dominated by the derivative terms, since $\mathcal{H}^{-1}\Phi'\gg \Phi$. With $x= k\eta$ and $\tilde{x}= k\tilde{\eta}$, we obtain \begin{equation} 
I_{\rm post}(u,v,k,\eta) \simeq \frac{uv}{2x} \int_{x_{\rm evap}}^x d\tilde{x} \tilde{x}^3 \sin(x-\tilde{x})  \frac{dT_{\Phi,{\rm post}}(vk,v\tilde{x})}{d(v\tilde{x})} \frac{dT_{\Phi,{\rm post}}(uk,u\tilde{x})}{d(u\tilde{x})}. \label{eq:Ipost_derivative} \end{equation} Substituting Eq.~\eqref{eq:dTpost_dx} and taking the late-time limit $x\gg x_{\rm evap}$ gives 
\begin{equation} \begin{aligned} I_{\rm post}(u,v,k,\eta) \simeq& \frac{c_s^2uv}{2x} x_{\rm evap}^4 \mathscr{S}_\Phi(vk)\mathscr{S}_\Phi(uk) T_{\Phi,{\rm eMD}}(vk)T_{\Phi,{\rm eMD}}(uk) \\ &\times \int_0^\infty \frac{d\tilde{x}}{\tilde{x}+x_{\rm evap}}  \sin(\bar{x}-\tilde{x})  \sin(c_s v\tilde{x}+\phi_{vk})  \sin(c_s u\tilde{x}+\phi_{uk}) , \end{aligned} \label{eq:Ipost_integral} \end{equation} where $\bar{x}= x-x_{\rm evap}$, and we shifted the integration variable by $\tilde{x}\to\tilde{x}+x_{\rm evap}$. Constant phase shifts generated by this change of variables have been absorbed into $\phi_{vk}$ and $\phi_{uk}$. 

The integral in Eq.~\eqref{eq:Ipost_integral} is dominated by the resonant configuration. To highlight this, we expand the product of sines into oscillatory terms. Contributions with nonzero net frequency are suppressed by rapid phase oscillations, while the slowly varying term arises when \begin{equation} 1-c_s(u+v)\simeq 0 . \end{equation} Equivalently, the tensor mode frequency is matched by the sum of the two scalar acoustic frequencies. 

Focusing on the resonant contribution, one obtains  
\begin{equation} 
\begin{aligned} \int_0^\infty \frac{d\tilde{x}}{\tilde{x}+x_{\rm evap}}  \sin(\bar{x}-\tilde{x})  \sin(c_s v\tilde{x}+\phi_{vk})  \sin(c_s u\tilde{x}+\phi_{uk}) \simeq&~ \frac{1}{4} {\rm Ci} \left(|1-c_s(u+v)|x_{\rm evap}\right) \\ &\times \sin\left(\bar{x}+|1-c_s(u+v)|x_{\rm evap}+ \phi_{vk}+\phi_{uk}\right) , \end{aligned} \label{eq:resonant_integral} \end{equation} 
up to terms that remain non-singular near $c_s(u+v)=1$. 
  After averaging over the rapid tensor oscillations, the phase dependence drops out. Retaining the resonantly enhanced contribution, which controls the high frequency scaling, the time averaged squared kernel is
  \begin{equation} \overline{I_{\rm post}^2} \simeq \frac{c_s^4u^2v^2}{2^7x^2} x_{\rm evap}^8 \mathscr{S}_\Phi^2(vk)\mathscr{S}_\Phi^2(uk) T_{\Phi,{\rm eMD}}^2(vk)T_{\Phi,{\rm eMD}}^2(uk) {\rm Ci}^2 \left(|1-c_s(u+v)|x_{\rm evap}\right). \label{eq:I2post} \end{equation}
  This expression captures the high frequency scaling of the post-evaporation contribution used in the main text.

\subsubsection{E3. Gravitational waves}

The contribution to the GW energy spectrum sourced during the post-evaporation epoch is then given by
\begin{eqnarray}
    \Omega_{\rm GW,post}(k)=\frac{2k^2}{3{\cal H}^2}\int_0^{\infty}dv\int_{|1-v|}^{1+v}du\left[\frac{4v^2-(1+v^2-u^2)^2}{4vu}\right]^2 \mathcal{P}_{\mathrm{S}}(vk)\mathcal{P}_{\mathrm{S}}(uk)\overline{I^2_{\rm post}(u,v,k,x)}~.
\end{eqnarray}
Substituting \eqref{eq:I2post}  into the above expression, we obtain
\begin{eqnarray}
\begin{aligned}
    \Omega_{\rm GW,post}(k)  &
    \simeq  \frac{c_s^4}{192}x_{\rm evap}^8\mathscr{S}_\Phi^4(k)T_{\Phi,{\rm eMD}}^4(k)\mathcal{P}_{\mathrm{S}}^2(k)\int_0^{\infty} dv\int_{|1-v|}^{1+v} du\left[\frac{4v^2-(1 + v^2 - u^2)^2}{4vu}\right]^2(vu)^{1 - 2b}\cr
    & ~~~~~~~~~~~~~~~~~~~~~~~~~~~~~~~~~~~~~~~~~~~~~~~~~~~~~~~~~~~~~~~~~~~\times{\rm Ci}^2\big(|1 - c_s(u + v)|x_{\rm evap}\big)~.
    \end{aligned}
\end{eqnarray}
For the integral above, we change variables according to
\begin{eqnarray}
    y&=&\big(c_s(u+v)-1\big)x_{\rm evap}~,\\
    s&=&u-v~,
\end{eqnarray}
where the Jacobian of the transformation is
\begin{eqnarray}
    |J|=\frac{1}{2c_s x_{\rm evap}}~.
\end{eqnarray}
Since in the frequency range of our interest, $x_{\rm evap}\gg1$ and ${\rm Ci}^2(|y|)$ is sharply peaked at $y=0$. We can approximate  $f(y,s){\rm Ci}^2(|y|)\to f(0,s){\rm Ci}^2(|y|)$. Consequently, we use
\begin{eqnarray}
    \int_{-\infty}^\infty {\rm Ci}^2(|y|)dy=\pi~
\end{eqnarray}
to approximate the integration over $y$ variable.  

The result can be recast as 
\begin{eqnarray}
    \Omega_{\rm GW,post}(k)\simeq\frac{c_s^{1+4b}(1-c_s^2)^2C^4(w){\cal A}^4_\Phi}{2^{5-4b}3^{3}\pi}\left(\frac{k}{k_{\rm evap}}\right)^{5-4b}\left(\frac{k_{\rm uv}}{k_{\rm evap}}\right)^2\left(\frac{k_{\rm d}}{k_{\rm uv}}\right)^8\Theta_{\rm uv}\left(k\right)~, 
\end{eqnarray}
where the function $\Theta_{\rm uv}$ comes from the remaining integral over $s$, which can be approximated as 
\begin{eqnarray}
\begin{aligned}
\Theta_{\rm uv}(k) \simeq \int_{-s_0(k)}^{s_0(k)} & \frac{(1-s^2)^2}{(1-c_s^2s^2)^{1+2b}} ds~~\text{with}~~s_0(k) = \max\left[ 0, \min\left( 1, 2\frac{k_{\rm uv}}{k} - c_s^{-1} \right) \right]~.
\end{aligned}
\end{eqnarray}

\end{document}